\documentclass[12pt]{article}
\usepackage{amsmath}
\usepackage{amsfonts}
\usepackage[dvips]{graphics,color}
\begin{document}
\parindent 0 pt



\def \ep{\epsilon} 
\def \intR {\int_{-\infty}^{+\infty}}  
\def \grad {\nabla}  
\def \ov{\over} 
\def \q  {\quad}  
\def \qq {\qquad}  
\def \bar{\overline} 
\def \beq{ \begin{equation} } 
\def \eeq{\end{equation}} 
\def \r#1{$^{#1}$}
\newtheorem{proposition}{Proposition}
\newtheorem{lemma}{Lemma}                 
\newtheorem{theorem}{Theorem}
\newtheorem{corollary}{Corollary}


\title{On the Global Dynamics of the Anisotropic Manev Problem}
\author{{Florin
Diacu}\renewcommand{\thefootnote}{\alph{footnote})}\footnotemark[1]~~
and{                                                            Manuele
Santoprete}\renewcommand{\thefootnote}{\alph{footnote})}\footnotemark[2]
\\ Department of Mathematics and Statistics \\ University of Victoria,
P.O.  Box 3045 Victoria B.C., \\ Canada, V8W 3P4}

\renewcommand{\thefootnote}{\alph{footnote})}

\maketitle   \footnotetext[1]{Electronic   mail:   diacu@math.uvic.ca}
\footnotetext[2]{Electronic mail: msantopr@math.uvic.ca}
\begin{abstract}

\noindent We study the global flow of the anisotropic Manev problem, which 
describes the planar motion of two bodies under the influence of an anisotropic 
Newtonian potential with a relativistic correction term. We first find all the 
heteroclinic orbits between equilibrium solutions. Then we generalize the Poincar\'e-Melnikov method and use it to prove the existence of infinitely 
many transversal homoclinic orbits. Invoking a variational principle and 
the symmetries of the system, we finally detect infinitely many classes of 
periodic solutions.

\end{abstract}

\small\normalsize
\section{Introduction}

The anisotropic Manev problem describes the motion of two point
masses in an anisotropic configuration plane under the influence of a 
Newtonian force-law with a relativistic correction term. The isotropic
case is the classical Manev problem; its origins lie in the work
of Newton, who introduced it in {\it Principia} aiming to understand 
the apsidal motion of the moon (see \cite{Delgado,Diacu01}). Manev
found in the 1930s that a proper choice of the constants that show
up in the correction term allow the theoretical explanation of the 
perihelion advance of Mercury and of the other inner planets.

The first author suggested the study of the anisotropic Manev problem 
in 1995, hoping to find connections between classical, quantum, and 
relativistic mechanics. It was indeed proved in \cite{Diacu1} that the rich collision-orbit manifold of the
system exhibits classical, quantum, and relativistic properties. This
encouraged further studies, as for example \cite{Diacu2} and \cite{Santoprete}.
In \cite{Diacu2}, using a suitable generalization of the Poincar\'e-Melnikov 
method (see \cite{Gallavotti,Holmes,Melnikov,Wiggins} for the classical 
approach or \cite{Chow,Chow1} for a parallel, at least in part, complementary 
approach), we proved that chaos occurs on the zero-energy manifold, thus
showing the complexity of the dynamics. Using perturbations techniques 
and the Poincar\'e continuation method, the second author investigated in 
\cite{Santoprete} the classes of periodic solutions that arise from 
symmetries in the case of small values of the anisotropy parameter. 

In this paper we gain a better understanding of the complicated
global dynamics encountered in this problem. We first prove that negative-energy 
solutions are bounded and find the heteroclinic orbits that connect the equilibria 
of the collision manifold, which we obtain through McGehee-type transformations
(see \cite{McGehee}). Physically they correspond to ejection-collision orbits. 
Then we employ perturbation techniques to detect possible global chaotic behaviour. 
As remarked in \cite{Santoprete}, the perturbation analysis of \cite{Diacu2,Santoprete} 
cannot be used to study ejection-collision solutions. However, we surpass this 
difficulty with the help of McGehee-type coordinates, which allow us to view 
the anisotropic Manev problem as a perturbation of the classical Manev case. 

Using an approach inspired by \cite{Chow,Chow1}, which works in some 
degenerate cases---as for example those of unstable nonhyperbolic 
points or critical points located at infinity (see \cite{Cicogna0,Cicogna,Cicogna1,Diacu2}), we 
develop a suitable extension of the Poincar\'e-Melnikov method, which we use to 
prove the existence of transversal homoclinic orbits to a periodic one. It is 
interesting to note that our result extends the one obtained in \cite{Cicogna0,Cicogna,Cicogna1} to a non-Hamiltonian system that has 
negatively and positively asymptotic sets to a nonhyperbolic 
periodic orbit. In the present context the asymptotic sets are the stable 
and the unstable manifolds. 

Then we return to the original coordinates and apply a variational
principle for detecting periodic orbits. Using the
rotation index, we divide the set of periodic paths into homotopy 
classes, which are Sobolev spaces. Then we use the lower
semicontinuity version of Hilbert's direct method (due to Tonelli,
see \cite{Tonelli}) to find a minimizer 
of the action in each class. According 
to the least action principle, the minimizer is a solution of the 
anisotropic Manev problem. We prove that the minimizer
exists, belongs to the homotopy class, and is a solution in the
classical sense. This generalizes a result obtained by the
second author, \cite{Santoprete}, where it was shown that 
such orbits exist for small values of $\mu>1$. In the end we 
put into the evidence some new properties of symmetric periodic orbits.
 
The idea of using variational principles to obtain periodic orbits 
for $n$-body-type particle systems first appeared in \cite{Poincare} 
and has been recently used in connection with symmetry conditions to 
obtain new periodic orbits in the classical $n$-body problem 
(see \cite{Chenciner}). But unlike the 
Newtonian case, the Manev force is ``strong'' (as defined in \cite{Gordon}), 
so the variational method is easier to apply in our situation than in
the Newtonian one. This is because in the Manev case we do not have to 
deal with the difficulty of avoiding collision orbits, which have infinite 
action and therefore cannot be minimizers. 
 
Our paper is organized as follows. In Section~2 we write the equations 
of motion and transform them to an equivalent system using a ``blow-up'' 
technique devised by McGehee, which allow us to introduce the concept 
of a collision manifold. In Section~3 we present  two  global results:  
the boundedness of the solutions for negative energy and the existence 
of certain symmetric ejection-collision orbits. In Section~4 we describe 
the anisotropic Manev problem as a perturbation of the Manev case. In 
Section~5 we develop a suitable generalization of the Poincar\'e-Melnikov   
method and in Section~6 we apply it to find infinitely many transverse 
homoclinic orbits that show that the dynamics of the problem is extremely 
complex, possibly chaotic. Finally, in Section~7 we use a variational 
principle to prove the existence of infinitely many classes of symmetric 
periodic orbits.

\section{Equations of Motion}

The (planar) anisotropic Manev problem is described by the Hamiltonian
\beq 
H={1 \over  2}{\bf p}^2 - {1 \over  \sqrt{x^2+ \mu y^2}}-{b \over
x^2 +\mu y^2},
\label{H}
\eeq 
where $\mu>1$  is a constant, ${\bf  q}=(x,y)$ is the position
of one body  with respect to the other considered  fixed at the origin
of the coordinate  system, and ${\bf p}=(p_x,p_y)$ is the momentum of
the moving particle.  The constant $\mu$ measures the strength of the
anisotropy and we can very well take $\mu<1$; but to remain consistent
with the choice made in previous papers, we will consider $\mu>1$. For 
$\mu=1$ we  recover the classical Manev problem.
The equations of motion are \beq \left\{
\begin{array}{l}
\dot  {\bf  q}  =  {\bf  p}  \\  \dot{\bf  p}  =  -{\partial  H  \over
\partial{\bf q}}.
\end{array}
\right. 
\label{eqmotion}
\eeq The Hamiltonian provides the first integral
\beq
H({\bf p}(t),{\bf q}(t))=h,
\label{energy}
\eeq
where $h$ is a real constant. Unlike in the classical Manev case, the 
angular momentum $K(t)={\bf p}(t)\times {\bf q}(t)$ does not yield 
a first integral. This is because the anisotropy of the plane destroys 
the rotational invariance.

Since our first goal is to study collision and near collision solutions, 
it is helpful to transform system (\ref{eqmotion}) using a method
developed by McGehee \cite{McGehee}. The idea is to ``blow-up'' the
collision singularity, replace it with a so-called collision manifold 
and extend the phase space to it. The collision manifold is fictitious
in the sense that it has no physical meaning. However, studying the flow 
on it provides useful information about near-collision orbits. Consider 
the coordinate transformations
\beq
\left\{
\begin{array}{l}
r=|{\bf q}|\\
\theta=\arctan(y/x)\\
v=\dot r r=(xp_x+yp_y)\\
u=r^2\dot \theta=(xp_y-yp_x),
\end{array}
\right.
\eeq
and the rescaling of time 
\beq
d\tau=r^{-2}dt.
\eeq
Composing these transformations, which are analytic diffeomorphisms in 
their respective domains, system
(\ref{eqmotion}) becomes
\beq   \left\{   \begin{array}{l}  r'=rv\\   v'=2r^2h+r\Delta^{-1/2}\\
\theta'=u\\ u'=(1/2)(\mu-1)(r\Delta^{-3/2}+2b\Delta^{-2})\sin 2\theta 
\label{McGehee}
\end{array} \right.
\eeq  and  the energy  relation  (\ref{energy})  takes  the form  \beq
u^2+v^2-2r\Delta^{-1/2}-2b\Delta^{-1}=2r^2h,  
\label{energyrelation}
\eeq where $\Delta=\mu\cos^2\theta+\sin^2\theta$ and the new variables
$(r,v,\theta,u)\in  (0,\infty)\times {\mathbb  R}\times  S^1\times{\mathbb R}$
depend on the fictitious time $\tau$. The prime denotes differentiation
with respect to $\tau$.

The set  \beq C= \{  (r,v,\theta,u)|r=0 \mbox{ and the  energy relation
(\ref{energyrelation}) holds} \} \eeq is the {\it collision manifold},
which replaces the set of singularities $\{ ({\bf q},{\bf p})|{\bf q=0}\}$.
This 2-dimensional manifold, embedded in ${\mathbb R}^3\times S^1$, is 
homeomorphic to a torus and it is given by the equations \beq r=0 \q 
\mbox{and} \q u^2+v^2=2b\Delta^{-1}.   \eeq
The flow on the collision manifold was studied in detail in \cite{Diacu1}. 
Here we will briefly recall its main features. Let's consider the restriction 
of system (\ref{McGehee}) to $C$. The solutions of the restriction lie on 
the level curves $v=\mbox{constant}$ of the torus $C$. There are eight  
equilibrium  points. In the variables $(r,v,\theta,u)$ the first four 
equilibria are $A_0^{\pm}=(0,\pm\sqrt{2b/\mu},0,0)$ and
$A_\pi^\pm=(0,\pm\sqrt{2b/\mu},\pi,0)$.  The corresponding eigenvalues
are real and take the values $\pm\sqrt{2b/\mu},0,\pm\sqrt{2b(1-\mu)/\mu}$.     
The other four equilibria are $A_{\pm \pi/2}^\pm=(0,\pm\sqrt{2b},\pm\pi/2,0)$  
and the corresponding eigenvalues are $\pm\sqrt{2b},0,\pm\sqrt{2b(1-\mu)}$,  
where the last two eigenvalues are purely imaginary since $\mu>1$.   
Moreover there are eight heteroclinic orbits which lie in the level sets
$v=\pm\sqrt{2b/\mu}$. All the other solutions are periodic (see Fig.~1).

\begin{figure}[h]
\begin{center}
\resizebox{!}{6.5cm}{\includegraphics{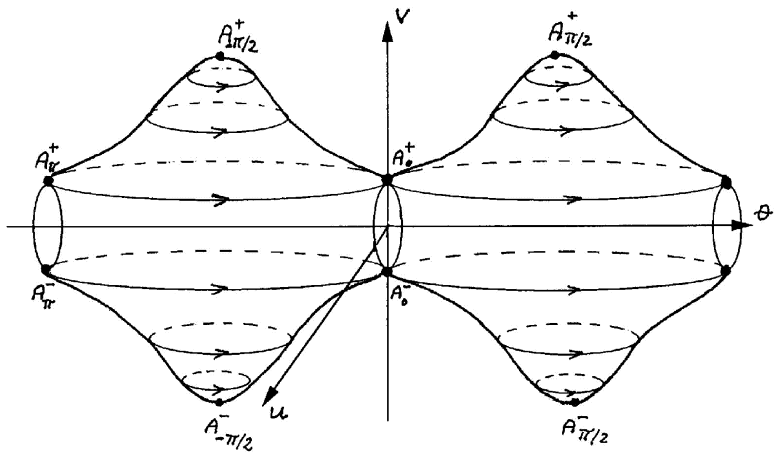}}
\end{center}
\caption[Fig. 1]{The flow on the collision manifold, which is formed by
periodic orbits, eight equilibria, and eight heteroclinic orbits.
}
\end{figure}

\section{Heteroclinic Orbits and Bounded Solutions}

The flow near the collision manifold was studied in \cite{Diacu1},  
in which most of the results are essentially local. In this section  
we will prove two global results that extend the understanding of 
the problem under discussion. The first one concerns the boundedness 
of solutions on negative energy levels.

\begin{theorem} For any negative value of the energy constant, $h<0$,  
there exists a positive real number $M$ such that any given solution 
$(r(\tau),v(\tau),\theta(\tau),u(\tau))$ of system (\ref{McGehee})
satisfies the relation $r<M$.
\end{theorem}

{\it Proof:}  Let us assume that there is no $M$ with the above property. 
Then at least one unbounded solution exists. Since by the energy relation (\ref{energyrelation}), $u^2+ v^2=2r^2h+2r\Delta^{-1/2}+2b\Delta^{-1}$, 
and since $h<0$, there is some $\bar r=r(\bar\tau)$ such that $u^2+v^2$ 
is negative---a contradiction. This completes the proof.

\smallskip

The next result deals with the existence of heteroclinic orbits
connecting the equilibria but lying outside the collision manifold. 
But before stating and proving it, let us recall some facts that 
summarize the behavior of the flow near the collision manifold. 
Denote by $P_\eta$ the periodic orbit on $C$ having $v=\eta$. 
The following property was proved in \cite{Diacu1}. 

\begin{proposition}
On the collision manifold $C$ the equilibria $A_0^\pm$ and $A_\pi^\pm$
are saddles whereas the equilibria $A_{\pm \pi/2}^\pm$ are centers.  
Outside the collision manifold the equilibria $A_0^\pm$, $A_{\pm \pi/2}^+$, 
and $A_\pi^+$ have a $1$-dimensional unstable analytic manifold, whereas the  
equilibria $A_0^-$,  $A_{\pm \pi/2}$, and $A_\pi^-$ have a $1$-dimensional 
stable analytic manifold. Each periodic orbit $P_\eta$ on $C$ with  
$v=\eta>0$($v=\eta<0$) has a $2$-dimensional local unstable analytic 
manifold, while the periodic orbit  $v=0$  has both a $2$-dimensional  
local unstable and a $2$-dimensional local stable manifold (see Fig.~2)
\end{proposition}

The above properties are local, the following one, however, is global. 
We will now show that the equilibria with positive $v$ coordinate have 
a 1-dimensional global unstable manifold while the equilibria with a 
negative $v$ have a 1-dimensional stable manifold. Moreover, the equilibria 
are connected by heteroclinic orbits starting from an equilibrium with positive 
$v$ and ending in the symmetric one with respect to the $(\theta,u)$ plane.

\begin{theorem}
There are four heteroclinic orbits outside the collision manifold $C$:
$\gamma_{-\pi/2}$ ,$ \gamma_0$ ,$ \gamma_{\pi/2}$ ,$ \gamma_{\pi}$
connecting respectively  $A_{-\pi/2}^+$ with $A_{-\pi/2}^-$, $A_0^+$ 
with $A_0^-$,  $A_{\pi/2}^+$ with $A_{\pi/2}^-$, and $A_\pi^+$ with 
$A_\pi^-$ (see Fig.~2)
\end{theorem}

{\it Proof:} First we show that $u=0$ and $\theta=0,\pi, \pm\pi/2$ describe 
four invariant sets. Consider $\theta(0)=\theta_0=0$ and $u(0)=u_0=0$, 
as initial conditions. Then $\theta\equiv 0$, $u\equiv 0$ satisfies  system
(\ref{McGehee}), hence (by the uniqueness property for solutions) $u=0$, 
$\theta=0$ define an invariant set. The same reasoning can be applied if 
$\theta=\pm \pi/2$ or $\pi$.

Now let's study the energy relation (\ref{energyrelation}) when $u=0$ and 
$\theta=0,\pi$. After simple computations we get
\beq
v^2+\left(\sqrt{2|h|}r-{1 \over \sqrt{2\mu|h|}}\right)^2={2b\over \mu}
+{1\over 2\mu|h|}.
\eeq
The above equation describes an ellipse whose intersections with  
$r=0$ give $v=\pm \sqrt{2b/\mu}$, which are exactly the equilibrium points 
$A_0^\pm$ and $A_\pi^\pm$. Moreover  the maximum value of $|v|$ is 
\beq
v_{max}=\sqrt{{1\over \mu}\Big(2b+{1\over 2|h|}\Big)}
\eeq
and the maximum value of $r$, attained when  $v=0$, is
\beq
r_{max}= {1\over (2\sqrt{\mu}|h|)}+{\sqrt{{1 \over \mu}(1-4hb)} \over 2|h|}.
\eeq
Consequently for $u=0$, $\theta=0$ ($\theta=\pi$) there exist heteroclinic orbits
$\gamma_0$, ($\gamma_\pi$) ejecting from $A_0^+$ ($A_0^-$) and tending to $A_0^-$ ($A_\pi^-$) (see Fig.~2).

Similarly when $u=0$ and $\theta=\pm \pi/2$ the energy relation can be reduced
to the form
\beq
v^2+\left(\sqrt{2|h|}r-{1 \over \sqrt{2|h|}}\right)^2=2b+{1 \over 2|h|},
\eeq
which describes an ellipse. The intersections with $r=0$ 
are $v=\pm\sqrt{2b}$ and represent the equilibria $A_{\pm \pi/2}^\pm$.
In this case 
\beq
v_{max}=\sqrt{2b+1/2|h|}
\eeq 
and 
\beq
r_{max}={1 \over 2|h|} +{\sqrt{1+4b|h|} \over 2|h|}.
\eeq
Thus we found heteroclinic orbits $\gamma_{\pm \pi/2}$ ejecting from $A_{\pm \pi/2}^+$ and tending to $A_{\pm\pi/2}^-$ (see Fig.~2). This completes the proof.

\begin{figure}[h]
\begin{center}
\resizebox{!}{7cm}{\includegraphics{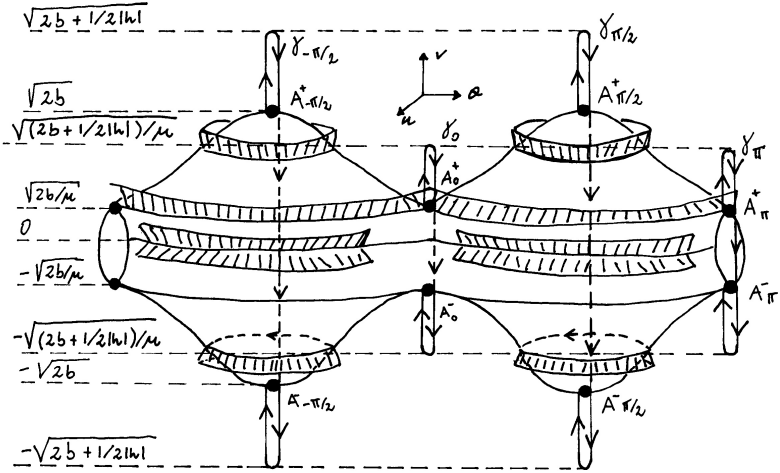}}
\end{center}
\caption[Fig. 2]{The flow can reach the collision manifold at the equilibria or at any of the periodic orbits. There are four heteroclinic orbits $\gamma_{-\pi/2}$ ,$ \gamma_0$ ,$ \gamma_{\pi/2}$ ,$ \gamma_{\pi}$ connecting respectively  $A_{-\pi/2}^+$ with $A_{-\pi/2}^-$, $A_0^+$ with $A_0^-$,  $A_{\pi/2}^+$ with $A_{\pi/2}^-$, and $A_\pi^+$ with $A_\pi^-$.}
\end{figure}

\section{A Perturbative Approach}

We will now write the anisotropic Manev problem as a perturbation of
the classical Manev case. Consider weak anisotropies, i.e., choose the  
parameter $\mu$ close to $1$. Introducing the notation $\mu-1=\epsilon>0$ 
with $\epsilon \ll 1$, we can expand the equation of motion in powers of
$\epsilon$ to obtain
\beq \left\{\begin{array}{l} r'=rv\\ v'=2r^2h+r
-\ep(r/2\cos^2\theta)\\ \theta'=u\\ u'=\ep(r/2+2b)\sin 2\theta.
\end{array} \right.
\label{perturb}
\eeq
The energy relation becomes  
\beq
u^2+v^2-2r-2b+\ep(r+2b)\cos^2\theta=2r^2h.
\label{enperturb}
\eeq   
For $\ep=0$, system (\ref{perturb}) and equation (\ref{enperturb})
yield the Manev problem. The collision manifold is the 
set of solutions given by
\beq 
r=0, \qq  u^2+v^2=2b.  
\eeq 
Notice that, from the geometric point of view, the collision manifold  
is a cylinder in the three-dimensional space of coordinates  $(u,\theta,v)$
and, since $\theta \in [0,2\pi]$, it follows that this cylinder can be 
identified with a torus. The flow on the collision manifold is formed  
almost exclusively by non-hyperbolic periodic orbits, except for the upper  
and lower circles of  the torus given by $r=0, \q u=0,\q  v=\pm \sqrt{2b}$,  
which consist of equilibrium  points. There is only a single orbit
ejecting from each fixed point of the upper circle $v=\sqrt{2b}$ and
a single orbit tending to the lower circle $v=-\sqrt{2b}$  (see
\cite{Diacu0}). Moreover it can be easily proved (see \cite{Diacu0}) 
that for every periodic orbit $p_v$ on the collision manifold  with 
$0<v<\sqrt{2b}$  there  exist a  manifold of orbits, lying on a cylinder,
which eject from $p_v$. Similarly it can be shown  that  for every  orbit  
$p_v$,  with $-\sqrt{2b}<v<0$,  there exists  a manifold of  orbits, lying  
on a cylinder, which  tend to $p_v$. 

If $v=0$ both types of manifolds exist, so $p_0$ has a homoclinic manifold. 
Indeed, the equations that describe the manifold can be found explicitly: 
they have $u=\pm \sqrt{2b}$. With the energy relation we get 
\beq 
v=\pm  \sqrt{2r^2h+2r}, 
\eeq  
and using  the equation of motion we obtain \beq r'=\pm r \sqrt{2r^2h+2r}.
\label{r'}
\eeq
By integrating  equation (\ref{r'}) it is easy  to find that
\beq R(\tau-\tau_0)={2
\over   2|h|+(\tau-\tau_0)^2}, \q   R'=-{4(\tau-\tau_0)\over
(2|h|+(\tau-\tau_0)^2)^2}  
\label{homman1}
\eeq    
and    
\beq    V(\tau-\tau_0)={R'    \over
R}=-{2(\tau-\tau_0)\over  2|h|+(\tau-\tau_0)^2}.  
\label{homman2}
\eeq  
Furthermore
\beq  U(\tau-\tau_0)=\pm  \sqrt{2b}=\omega  \q \mbox{and}  \q
\vartheta(\tau-\tau_0,\theta_0)=\Theta(\tau-\tau_0)-\theta_0, 
\label{homman3}
\eeq    
where $\Theta(\tau-\tau_0)=\omega(\tau-\tau_0)$. As $\tau_0$ 
and $\theta_0$ vary, equations (\ref{homman1}-\ref{homman3}) 
describe the entire $2$-dimensional homoclinic manifold.
An orbit lying on the homoclinic manifold  is represented in 
Fig.~3. Such an orbit is obtained by choosing $\theta_0=0$; 
it ejects from the equator of the collision manifold, spiraling 
around it and moving upwards, then changes directions, 
goes downwards and upwards again, spiraling towards the 
periodic orbit $p_0$. 
\begin{figure}[h]
\begin{center}
\resizebox{!}{5cm}{\includegraphics{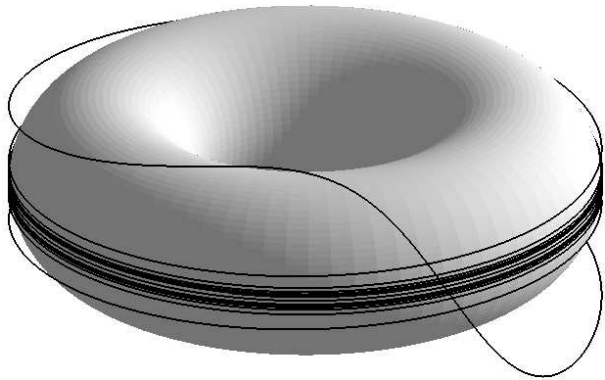}}
\end{center}
\caption[Fig. 3]{An homoclinic orbit to $p_0$ lying on the homoclinic 
manifold. This orbit spirals out of the equator of the collision manifold 
and then spirals back to it.}
\end{figure}

The homoclinic manifold plays an important role in following section and 
is necessary for developing the generalization of the Melnikov technique.

\section{A Generalized Melnikov Method}

Let $\chi=(R(\tau),V(\tau),\Theta(\tau),U(\tau))$ be the homoclinic orbit 
selected when we choose $\tau_0=0$ and $\theta_0=0$. Consider solutions 
of the form
\beq \left\{
\begin{array}{l}
r(\tau,\tau_0)=R(\tau-\tau_0)+  \tilde r(\tau,\tau_0)\\
 v(\tau,\tau_0)=V(\tau-\tau_0)+ \tilde v(\tau,\tau_0)\\
\theta(\tau,\tau_0,\theta_0)=\Theta(\tau-\tau_0)-\theta_0 + \tilde \theta(\tau,\tau_0)\\ 
u(\tau,\tau_0)=U(\tau-\tau_0)+ \tilde u(\tau,\tau_0).
\end{array}
\right.  \eeq 
Let ${\bf\tilde z}=(\tilde r,\tilde v,\tilde \theta, \tilde u)$, then the variational
equation is
\beq  {\bf
\tilde z}'=A(\tau){\bf\tilde z}+{\bf \tilde  b}({\bf \tilde z}, \chi,\tau,\tau_0,\theta_0,\epsilon),
\label{variational}
\eeq  
where
\beq 
A(\tau)=
  \left(
\begin{array}{cccc}
V & R & 0 & 0\\ 1+4Rh & 0 & 0 &  0\\ 0 & 0 & 0 & 1\\ 0 & 0 & 0 &
0
\end{array}\right) 
\eeq
and 
\beq
{\bf \tilde b}({\bf \tilde z},\chi,\tau,\tau_0,\theta_0,\epsilon)=
\left( \begin{array}{c}
b_1\\
b_2\\
b_3\\
b_4
\end{array}
\right)
=\left (
\begin{array}{c}
\tilde r(\tau-\tau_0) \tilde v(\tau-\tau_0) \\
 -\epsilon \left ({(R+\tilde r)\over 2} \cos^2(\Theta-\theta_0+\tilde \theta) \right )\\
0\\
\epsilon \left ( {(R+\tilde r)\over 2}+2b \right ) \sin 2(\Theta-\theta_0+\tilde\theta)
\end{array}
\right).
\eeq
The general solution of the variational equation (\ref{variational}) is
\beq 
{\bf\tilde  z}=\Phi(t)\int_{t_0}^{t} \Phi^{-1}(s){\bf\tilde b}~ ds, 
\eeq 
(see \cite{Hartman}), where $\Phi$ is the fundamental matrix. If we let $c=\Phi^{-1}{\bf\tilde b}$, the previous equation becomes          
\beq
\tilde z_i(t)=\Phi_{ij}\int_{t_0}^{t}c_j(s)~ds, 
\label{solution}\eeq           
where $c_j=det~D_j(t)/(det\Phi)(t)$ and $D_j$ is the matrix obtained
replacing the $j$-th column of $\Phi$ with ${\bf \tilde b}$. 
Furthermore the following formula for the trace holds: 
\beq
\mbox{det}\Phi(\tau)=Ce^{\int_{\tau_0}^\tau TrA(s)~ds}.
\label{trace}
\eeq  
One solution of the homogeneous part of the variational equation is
given by 
\beq
\chi'(\tau-\tau_0,\theta_0) =(R'(\tau-\tau_0),V'(\tau-\tau_0),\Theta'(\tau-\tau_0),U'(\tau-\tau_0)), 
\eeq
where 
\beq
\begin{array}{l}
R'=-{4(\tau-\tau_0)  \over (2|h|+(\tau-\tau_0)^2)^2}  \smallskip \\
V'=-{2   \over   2|h|+(\tau-\tau_0)^2}+{4(\tau-\tau_0)^2   \over
(2|h|+(\tau-\tau_0)^2)^2}\smallskip\\
\Theta'=\pm\sqrt{2b}.\smallskip\\
U'=0.
\end{array}
\eeq 
It is easy to check that other two independent solutions are $(0,0,1,0)$ and  $(0,0,0,1)$.  Knowing three independent solutions of a linear system, 
it is possible to find a fourth independent  solution $\psi$. This is 
achieved through the following lemma, which will be used to estimate how 
fast $\psi$ diverges.
\begin{lemma}Let ${\bf \tilde z}'=A{\bf \tilde z}$ be the homogeneous part of
(\ref{variational}). Given the three independent solutions above, a fourth 
is defined by
\beq
\left\{
\begin{array}{l}
\tilde z_1'=(1+4Rh)/V' \tilde z_4\smallskip\\
\tilde z_2'=\mp {\sqrt{2b}\over V'}(1+4Rh)\tilde z_4\smallskip\\
\tilde z_3'=0\smallskip\\
\tilde z_4'=(V-{R'(1+4Rh) \over V'}) \tilde z_4\\
\end{array}\right. \qq
\left \{
\begin{array}{l}
\psi_1=R'\tilde z_1+\tilde z_4\smallskip \\
\psi_2=V'\tilde z_1\smallskip\\
\psi_3=\pm\sqrt{2b}\tilde z_1 +\tilde z_2\\
\psi_4=\tilde z_3.
\end{array} \right.
\eeq
\end{lemma}
{\it Proof:} To find the fourth independent solution, we can use the
``reduction to a smaller system,''  (see \cite{Hartman}), whose direct
application completes the proof.

\smallskip

In  particular  it  is  useful  to remark that  we  can  always  choose
$\psi_4=0$, since  $(\psi_1,\psi_2,\psi_3,0)$ is a solution that is
independent from the other ones. To obtain necessary and sufficient
conditions such that the negatively and positively asymptotic sets 
intersect transversely, we first obtain conditions for the existence
of solutions bounded  on $\mathbb{R}$ for the  non-homogeneous linear
variational  equation  around  $\chi$.   

For this, let  ${\mathcal  B}({\mathbb
R})=\{{\bf\tilde  b}:{\mathbb R}\rightarrow  \mathbb{R  \times R}\times
S^1\times\mathbb{R}~\mbox{bounded, continuous}\}$  with $\|{\bf\tilde
b}\|=\mbox{sup}_{\tau\in{\mathbb R}}\|{\bf\tilde b}(\tau)\|$ for ${\bf
\tilde b}\in \cal{B}({\mathbb R})$. Then we have the following version
of the Fredholm alternative for solutions bounded on $\mathbb{R}$ (see \cite{Chow,Chow1,Cicogna}
for a similar approach).
\begin{lemma}
Let ${\bf \tilde  z}\in
\mathbb{R\times R\times} S^1\times\mathbb{R}$ and assume that   ${\bf\tilde z}\equiv0$  in the expression of the function ${\bf\tilde b}$.
Then the variational equation 
\beq
{\bf\tilde z}'=A(\tau){\bf\tilde z}+{\bf
\tilde  b}({\bf  \tilde z},  \chi,\tau,\tau_0,\theta_0,\epsilon)\q
\label{linearvariational}
\eeq 
has a bounded solution if and only if 
\beq \intR
e^{-\int_{\tau_0}^{\tau}  Tr A(s)ds}~R'(\tau-\tau_0)~b_2(\chi,\tau,\tau_0,\theta_0,\epsilon)~  d\tau=0.
\label{condition1} 
\eeq 
The solution is unique and continuous and has
the form ${\bf \tilde z}={\bf  L}({\bf \tilde b})+{\bf w}$, where ${\bf L}$ is
a bounded linear operator, ${\bf w}=(0,0, \tilde\theta(\tau_0),\tilde u(\infty))$,
when  $\tilde r(\tau_0)R'(\tau_0)+\tilde v(\tau_0)V'(\tau_0))=0$,
and $b_4$ satisfies the relation below,
\beq
\intR b_4(\chi,\tau,\tau_0,\theta_0,\epsilon)~d\tau=0.
\label{condition2}
\eeq 
\end{lemma}
{\it Proof:} Using Lemma~1 it is easy to determine  the behavior of $\psi$ as 
$\tau \rightarrow \pm \infty$, precisely,
\beq
\tau\rightarrow\pm\infty
\left \{
\begin{array}{l}
\tilde z_1\sim \tau^4\\
\tilde z_2\sim \tau^4\\
\tilde z_3\sim \mbox{const.}\\
\tilde z_4\sim \tau,\\
\end{array}
\right.
\q\q
\tau\rightarrow\pm\infty
\left \{
\begin{array}{l}
\psi_1\sim \tau\\
\psi_2\sim \tau^2\\
\psi_3\sim \tau^4\\
\psi_4\sim \mbox{const.}\\
\end{array}
\right.
\eeq
Using (\ref{solution}) and (\ref{trace}), the general solution of the complete (non-homogeneous) equation (\ref{variational}) can be written in integral form as
\beq
\begin{split}
\tilde r&=R' \left(A-\int_{\tau_0}^\tau e^{-\int_{s_0}^sTrA(\eta)d\eta}(\psi_1b_2-\psi_2b_1)~ds\right)\\
          &+\psi_1 \left(B+\int_{\tau_0}^\tau e^{-\int_{s_0}^sTrA(\eta)d\eta}(R'b_2-V'b_1)~ds\right)\\
\tilde v&=V' \left(A-\int_{\tau_0}^\tau e^{-\int_{s_0}^sTrA(\eta)d\eta}(\psi_1b_2-\psi_2b_1)~ds\right)\\
       &+\psi_2 \left(B+\int_{\tau_0}^\tau e^{-\int_{s_0}^sTrA(\eta)d\eta}(R'b_2-V'b_1)~ds\right)\\
\tilde\theta&=\pm\sqrt{2b}\left(A-\int_{\tau_0}^\tau e^{-\int_{s_0}^sTrA(\eta)d\eta}(\psi_1b_2-\psi_2b_1)~ds\right)\\
    &+\psi_3 \left(B+\int_{\tau_0}^\tau e^{-\int_{s_0}^sTrA(\eta)d\eta}(R'b_2-V'b_1)~ds\right)\\
     &+ C-\int_{\tau_0}^\tau e^{-\int_{s_0}^sTrA(\eta)d\eta} \left[(-V'\psi_3\pm\sqrt{2b}\psi_2)b_1+(R'\psi_3\pm\sqrt{2b}\psi_1)b_2\right]\\
\tilde u&=~D+\int_{\tau_0}^\tau b_4~ds,
\end{split}
\label{longsolution}
\eeq
where, for notational convenience, we failed to mention the dependence on 
${\bf \tilde z}$, $\chi$, $\tau_0$, etc.
 
Consider now the linearization of the problem (\ref{longsolution}) around the    
solution ${\bf\tilde z(\tau)}\equiv0$; in particular this amounts to deleting  
the high-order terms in the expression of ${\bf\tilde b}$ (i.e. $b_1=0$, etc.).
Taking  also  into  account  the
different behavior  of the different  solutions given in Lemma~1, it is
easy to  see that to  have bounded solutions  we need to  require that
\beq \psi_i\left(A-\int_{\tau_0}^\tau
e^{-\int_{s_0}^sTrA(\eta)d\eta}\psi_1b_2(\chi,s,\tau_0,\theta_0,\epsilon)~ds\right) 
\qq \mbox{for}~ i=1,\dots,4, 
\eeq  
remains bounded   as  $\tau\rightarrow  \pm\infty$. More precisely ${\bf\tilde z}$ 
is bounded on $[\tau_0,\infty)$ if and only if 
\beq A=\int_{\tau_0}^\infty
e^{-\int_{s_0}^sTrA(\eta)d\eta}\psi_1b_2~ds 
\label{semiA}
\eeq  
and bounded on $(-\infty,\tau_0]$ if and only if
\beq A=-\int_{-\infty}^{\tau_0}
e^{-\int_{s_0}^sTrA(\eta)d\eta}\psi_1b_2~ds. 
\label{semiA1}
\eeq  
We also require 
\beq
\tilde u(\pm\infty)=\lim_{\tau\to\ \pm\infty}\tilde u(\tau)=
D+\lim_{\tau\to \pm\infty}\int_{\tau_0}^\tau b_4(\chi,s,\tau_0,\theta_0,\epsilon)~ds,
\label{condition2}
\eeq
where, obviously, $\tilde u(\infty)=\tilde u(-\infty)$. 
The latter condition is not needed for the boundedness of the solution, but its role
will be clear later when analyzing some properties of the negatively and positively asymptotic sets. It  is  easy  to see  that  the  above
conditions are simultaneously satisfied  both at $\tau=-\infty$ and at
$\tau=+\infty$  if  for  some  $\tau_0$  the  following  Melnikov-type
conditions:
\beq
\begin{split}
&\intR e^{- \int_{\tau_0}^\tau TrA(\eta)~d\eta}\psi_1 b_2(\chi,s,\tau_0.\theta_0,\epsilon)~ds=0\\
&\intR b_4(\chi,s,\tau_0,\theta_0,\epsilon)~ds=0
\label{melnikovlike}
\end{split}
\eeq
are fulfilled. Thus we can rewrite the general solution (\ref{longsolution}) using
(\ref{melnikovlike}) and,  by neglecting to mention the dependence on $\chi$, $s$, etc., we obtain
\beq
\begin{split}
\tilde r&= -R'\int_{\infty}^\tau e^{-\int_{s_0}^sTrA(\eta)d\eta}\psi_1b_2~ds
          +\psi_1 \left(B+\int_{\tau_0}^\tau e^{-\int_{s_0}^sTrA(\eta)d\eta}R'b_2~ds\right)\\
\tilde v&=-V'\int_{\infty}^\tau e^{-\int_{s_0}^sTrA(\eta)d\eta}\psi_1b_2~ds
       +\psi_2 \left(B+\int_{\tau_0}^\tau e^{-\int_{s_0}^sTrA(\eta)d\eta}R'b_2~ds\right)\\
\tilde\theta&=\mp\sqrt{2b}\int_{\infty}^\tau e^{-\int_{s_0}^sTrA(\eta)d\eta}\psi_1b_2~ds
    +\psi_3 \left(B+\int_{\tau_0}^\tau e^{-\int_{s_0}^sTrA(\eta)d\eta}R'b_2~ds\right)\\
     &+ C-\int_{\tau_0}^\tau e^{-\int_{s_0}^sTrA(\eta)d\eta} (R'\psi_3\pm\sqrt{2b}\psi_1)b_2\\
\tilde u&=~\tilde u({\infty})+\int_{\infty}^\tau b_4~ds.
\end{split}
\label{longsolution2}
\eeq    
To obtain $\tilde r(\tau_0)R'(\tau_0)+\tilde v(\tau_0)V'(\tau_0))=0$ we must have
\beq
B={(R'^2(\tau_0)+V'^2(\tau_0)) \over \psi_1(\tau_0)R'(\tau_0)+\psi_2(\tau_0)V'(\tau_0)}
\int_\infty^{\tau_0} \psi_1(s)b_2(s)~ds.
\eeq 
Moreover we also get
\beq
C=\tilde\theta(\tau_0) \pm\sqrt{2b}\int_{\infty}^{\tau_0} e^{-\int_{s_0}^sTrA(\eta)d\eta}\psi_1b_2~ds\\
\eeq
and
\beq
D=\tilde u(\infty)+\int_\infty^{\tau_0} b_4~ds.
\eeq
This uniquely defines $B$, $C-\tilde\theta(\tau_0)$, and $D-\tilde u(\infty)$ as
continuous linear functionals on ${\cal B}(\mathbb{R})$. From (\ref{longsolution2}) we
observe that the corresponding solution is of the form ${\bf\tilde z}={\bf L}({\bf\tilde b})+{\bf w}$, where ${\bf L}$ is a bounded linear operator. It follows that this operator is continuous and hence the solution ${\bf\tilde z}={\bf L}({\bf\tilde b})+{\bf w}$ is continuous on ${\cal B}(\mathbb{R})$. This completes the proof.

\smallskip

To obtain necessary and sufficient conditions that the negatively and positively  
asymptotic sets intersect, let us first consider 
all the solutions of (\ref{variational}) which are bounded as $\tau \rightarrow 
-\infty$ and such that their angles remain close to the ones on the periodic orbit.
The solution ${\bf \tilde z}$ is given by (\ref{longsolution}) satisfying 
(\ref{semiA1}) and (\ref{condition2}) with negative sign. In particular the 
solutions of the variational equation that are bounded as $\tau \rightarrow 
-\infty$ (i.e. which remain in a sufficiently small neighborhood of the periodic 
orbit as $\tau \rightarrow -\infty$)  and with perturbed angles that do not 
drift but remain near the angles on the periodic orbit, must be on the negatively asymptotic set. In the same way, we obtain the positively invariant set from the 
solutions that remain bounded as $\tau \rightarrow \infty$ and whose angles stay 
close to the one of the periodic orbit, which was in fact the reason why we required 
that condition (\ref{condition2}) be satisfied.

 Moreover it is important to remark that the solution
 we found are not only bounded but also  such that $\tilde r \rightarrow 0$, $\tilde v\rightarrow 0$ as $\tau\rightarrow \infty$ and this is important since, on the collision manifold we have many periodic orbit and this condition is needed to show that the orbits are actually asymptotic to the equator.

With the preparations above, we can now prove the following result.
\begin{theorem}
System (\ref{perturb}) has transversal homoclinic solutions if and only if
there exist $\tau_0^*$ and a $\theta_0^*$ such that
\beq
\tilde M_1(\tau_0^*,\theta_0^*)=\tilde M_2(\tau_0^*,\theta_0^*)=0 \q
\mbox{and}\q {\partial \tilde M_1 \over \partial \tau_0}{\partial \tilde M_2 \over \partial \theta_0}
-{\partial \tilde M_1 \over \partial \theta_0}{\partial \tilde M_2 \over \partial \tau_0}
\left|_{\substack{\tau_0=\tau_0^*\\ \theta_0=\theta_0^*}}\right.\neq 0,
\eeq
where
\beq\begin{split}
\tilde M_1(\tau_0,\theta_0)&=\intR e^{-\int_{\tau_0}^{\tau}  Tr A(s)ds}~R'
~b_2({\bf \tilde z}^*,\tau,\tau_0,\theta_0,\epsilon)~  d\tau,\\ 
\tilde M_2(\tau_0,\theta_0)&=\intR b_4({\bf \tilde z}^*,\tau,\tau_0,\theta_0,\epsilon)~d\tau, 
\end{split}
\label{finalconditions}
\eeq
and ${\bf \tilde z}^*$ is a solution of ${\bf \tilde z}={\bf L}({\bf \tilde b}({\bf
 \tilde z},\tau,\tau_0,\theta_0,\ep))+{\bf  w}$.
Moreover if the perturbation is periodic we get infinitely many intersections.
\end{theorem}
{\it  Proof:} The stable and unstable manifolds intersect if and only if
the solution (\ref{longsolution}) satisfies the Melnikov-like conditions 
(\ref{condition1}) and (\ref{condition2}) of Lemma~2. This was already
proved in the case when ${\bf \tilde  b}$ did not implicitly depend 
depend on  ${\bf \tilde z}$. But because of this implicit dependence  
we need to apply the implicit function theorem, which states that given 
${\bf \tilde z}={\bf L}({\bf \tilde b}({\bf \tilde z},\tau,\tau_0,\theta_0,\ep))+{\bf  w}$ with ${\bf\tilde z
-w}={\bf  L}(0,\tau,\tau_0,\theta_0,0)=0$, there exist a $\delta$
and a unique solution ${\bf\tilde z^*(\epsilon,\tau_0,\theta_0)}$
(that has continuous derivatives up to order 2 in
$\tau_0,\theta_0,\epsilon$) such that $\epsilon<0$, $|{\bf\tilde
z}|<\delta$ if  the linearized operator ${\bf \tilde z}={\bf L}({\bf
\tilde b}(0,\tau,\tau_0,\theta_0,\ep))+{\bf w}$ is invertible.   
But Lemma~2 proved that such an operator is invertible. Moreover 
the homoclinic solutions are transversal if and only if the integrals (\ref{finalconditions}) have simple zeroes, as functions of $\tau_0$ 
and $\theta_0$ (see \cite{Chow,Chow1}). This concludes the proof.

\medskip

Unfortunately the Melnikov integrals of Theorem~3 are difficult
to compute explicitly. To overcome this difficulty we need to
rewrite these integrals to the first order approximation in 
$\epsilon$. Hence if we let ${\bf \tilde z}^*=\epsilon\chi$
and ${\bf \tilde b}=\epsilon {\bf d}$ with ${\bf d}=(d_1,d_2,d_3,d_4)$, 
the next result follows immediately.
\begin{corollary}
System (\ref{perturb}) has transversal homoclinic solutions if and only if
there exist $\tau_0^*$ and a $\theta_0^*$ such that
\beq
 M_1(\tau_0^*,    \theta_0^*)= M_2(\tau_0^*,    \theta_0^*)=0 \q
\mbox{and}\q {\partial  M_1 \over \partial \tau_0}{\partial  M_2 \over \partial     \theta_0}
-{\partial  M_1 \over \partial     \theta_0}{\partial  M_2 \over \partial \tau_0}
\left|_{\substack{\tau_0=\tau_0^*\\     \theta_0=    \theta_0^*}}\right.\neq 0,
\eeq
where
\beq\begin{split}
 M_1(\tau_0,\theta_0)&=\intR e^{-\int_{\tau_0}^{\tau}  Tr A(s)ds}~R'(\tau-\tau_0)
~b_2(\chi(\tau-\tau_0),\Theta(\tau-\tau_0)-\theta_0)~  d\tau,\\ 
 M_2(\tau_0,\theta_0)&=\intR b_4(\chi(\tau-\tau_0),\Theta(\tau-\tau_0)-\theta_0)~d\tau. 
\end{split}
\label{Mel}
\eeq
Moreover if the perturbation is periodic we get infinitely many intersections.
\end{corollary}
Corollary~1 generalizes the Melnikov integrals obtained in 
\cite{Holmes1,Wiggins} to nonhyperbolic whiskered tori (periodic orbits) in
non-Hamiltonian systems. We remark that the second integral in (\ref{Mel})
converges only conditionally. This is not a new feature  of this non-Hamiltonian 
system since  the same nuisance was present in \cite{Holmes1,Wiggins}.
However some authors, more recently,  found a way to write the Melnikov conditions for hyperbolic whiskered tori in Hamiltonian systems using only convergent integrals see \cite{Delshams,Treschev}. It would be interesting to generalize those results to nonhyperbolic tori in non-Hamiltonian systems and to apply the newly developed technique to the problem under discussion in this paper. But this is not a project we aim to develop here.

\section{The Melnikov Integrals}

Now we would like to apply Corollary~1 to our problem. The Melnikov 
conditions  take the form
\beq\begin{split}
 M_1(\tau_0,    {\theta}_0)=\intR \Big [e^{-{1 \over 2}\int_{\tau_0}^{\tau}V (s)ds}
R(\tau-\tau_0)R'(\tau-\tau_0) \\
\times  \cos^2(\omega(\tau-\tau_0)-    {\theta}_0) \Big ] ~
d\tau=0
\end{split}
\eeq 
  and  \beq  M_2(\tau_0,    {\theta}_0)={1\over  2}\intR
(R(\tau-\tau_0)+2b)\sin(2(\omega(\tau-\tau_0)-    {\theta}_0))~
d\tau=0.  \eeq Let  $\tilde\theta_0=-{\theta}_0-\omega\tau_0$. With
this assumption we can rewrite the first  Melnikov condition as 
\beq
M_1=\cos^2\tilde\theta_0I_1^a+\sin^2\tilde\theta_0I_1^b-\sin2\theta_0
I_1^c, 
\eeq 
where \beq \left\{
\begin{array}{l}
I_1^a=\intR    e^{-{1\over     2}\int_{\tau_0}^{\tau}    V(s)    ds}
RR'\cos^2\omega\tau~   d\tau\medskip\\  I_1^b=\intR  e^{-{1\over
2}\int_{\tau_0}^{\tau}     V(s)     ds}     RR'\sin^2\omega\tau~
d\tau\medskip\\   I_1^c=\intR   e^{-{1\over   2}\int_{\tau_0}^{\tau}
V(s) ds} RR'\sin\omega\tau\cos\omega\tau~ d\tau.
\end{array}
\right .  \eeq The second  Melnikov condition can be expressed as 
\beq
M_2=\cos2\tilde\theta_0I_2^a+\sin2\tilde\theta_0I_b^2,  
\eeq 
where 
\beq
\left \{
\begin{array}{l}
I_2^a={1\over2}\intR    (R+2b)\sin2\omega\tau~    d\tau   \medskip\\
I_2^b={1\over2}\intR (R+2b)\cos2\omega\tau~ d\tau. \\
\end{array}
\right .  \eeq
All  the  integrals  above  can  be computed  using  the  method  of
residues. Straightforward computations give \beq I_1^a=-I_1^b={-1 \over |h|}
\intR  {(\tau-\tau_0)\cos2\omega\tau  \over (2|h|+(\tau-\tau_0)^2)^2}~
d\tau=     {\pi\sin(2\omega\tau_0)e^{-2\omega    \sqrt{2|h|}}    \over
|h|\sqrt{2|h|}}   \eeq   and   \beq   I_1^c={-1   \over   |h|}   \intR
{(\tau-\tau_0)\sin\omega\tau\cos\omega\tau                        \over
(2|h|+(\tau-\tau_0)^2)^2}~                                       d\tau=
-{\pi\cos(2\omega\tau_0)e^{-2\omega\sqrt{2|h|}}\over   |h|\sqrt{2|h|}}.
\eeq 
Particular  care is needed  when integrating $I_2^a$  and $I_2^b$
since they converge only conditionally.  To obtain computational convergence,
we choose the limits in $I_2^a$ such that  
\beq\begin{split}
I_2^a&=\lim_{N\to\infty}\int_{-N\pi/2\omega}^{N\pi/2\omega}\left(b+{1
\over                2|h|+(\tau-\tau_0)^2}\right)\sin2\omega\tau~d\tau\\
&={\pi\sin(2\omega\tau_0)e^{-2\omega\sqrt{2|h|}}\over       \sqrt{2|h|}}.
\end{split}
\eeq 
The integral was also computed using the method of residues.
Similarly, for $I_2^b$, we have        
\beq\begin{split}
I_2^b& = \lim_{N\to\infty}  \int_{-N\pi/2\omega}^{N\pi/2\omega}\left(b+{1
\over2|h|+(\tau-\tau_0)^2}\right)\cos2\omega\tau~d\tau \\
 &= {\pi\cos(2\omega\tau_0)e^{-2\omega\sqrt{2|h|}}\over  \sqrt{2|h|}}
\end{split}
\eeq
and   thus   
\beq  M_1=M_2=   \sin(2(\omega\tau_0+\tilde\theta_0)){\pi
e^{-2\omega\sqrt{2|h|}}\over \sqrt{2|h|}}.
\eeq  
We therefore have only one independent condition; this is clearly a 
consequence of the energy relation.

We can find simple zeroes when $\sin(2(\omega\tau_0+\tilde\theta_0))=0$,
i.e., for $-(\omega\tau_0+\tilde\theta_0)= \theta_0=\pm k\pi/2$ for
$k=0,1,2,\dots$.

Hence, by Corollary 1, we have proved the existence of an infinite sequence
of  intersections on  the Poincar\'e  section of  the  negatively and
positively asymptotic sets of the periodic orbit and the existence of
homoclinic orbits  leaving the equator of the  collision manifold and
going  back to  it.  This  situation  is clearly  reminiscent of the
chaotic dynamics described by the Poincar\'e-Birkhoff-Smale theorem in 
terms of symbolic dynamics and the Smale horseshoe. Unfortunately
this theorem cannot be directly  applied, nor can the theorems proved
in \cite{Burns}, since the Poincar\'e-Birkhoff-Smale theorem considers 
hyperbolic fixed  points while the arguments in \cite{Burns} apply to
area-preserving diffeomorphisms.  However  the arguments contained in
those theorems strongly suggest the occurrence of a chaotic dynamics.

Moreover it is easy to verify, and interesting to remark, that the
orbits we found above are  not   $\bar  S_0$-symmetric, where  
the $\bar S_0$  symmetry is defined by $\bar   S_0(r,v,\theta,u,\tau)=(r,-v,-u,-\tau)$ (see \cite{Diacu1}) and  
an  orbit  $\gamma(\tau)$  is  said  to  be  $\bar
S_0$-symmetric   if  $\bar  S_0(\gamma(\tau))=\gamma(\tau)$. Indeed an orbit 
is $\bar S_0$-symmetric if and only if it has a point on the zero velocity curve, i.e., if there is a $\bar \tau$ such that $v(\bar\tau)= u(\bar\tau)=0$ (see \cite{Santoprete}). 
But this cannot happen in our problem because the unperturbed solution 
verifies $u\equiv\pm\sqrt{2b}$. Thus for $\epsilon$ small enough the 
perturbed orbit can never have $u=0$. 

\medskip
  
We can now summarize the above discussion as follows:

\begin{theorem}
Let us consider the anisotropic Manev problem given by the equation of
motion      (\ref{McGehee})     with      the      energy     relation
(\ref{energyrelation}).   Then   there  is  an   infinite  sequence  of
intersections  in  the  Poincar\'e   section  of  the  negatively  and
positively asymptotic  sets of the  periodic orbits at the  equator of
the   collision  manifold   (possibly   giving  rise   to  a   chaotic
dynamics).  Furthermore there  exist the  homoclinic non $\bar S_0$-symmetric 
  orbits to  the periodic orbit described above. 
\end{theorem}

\section{Periodic Solutions}

We now return to the original Cartesian coordinates, which
are more convenient for the purpose of finding certain periodic
solutions. Let us first notice that the equations (\ref{eqmotion})
admit the following symmetries:
$$
\begin{array}{l}
S_0(x,y,p_x,p_y,t)=(x,y,-p_x,-p_y,-t),\\
S_1(x,y,p_x,p_y,t)=(x,-y,-p_x,p_y,-t),\\
S_2(x,y,p_x,p_y,t)=(-x,y,p_x,-p_y,-t),\\
S_3(x,y,p_x,p_y,t)=(-x,-y,-p_x,-p_y,t),\\
S_4(x,y,p_x,p_y,t)=(-x,y,-p_x,p_y,t),\\
S_5(x,y,p_x,p_y,t)=(x,-y,p_x,-p_y,t),\\
S_6(x,y,p_x,p_y,t)=(-x,-y,p_x,p_y,-t),
\end{array}
$$
which are the elements of an Abelian group of order eight, isomorphic
to ${\bf Z}_2\times{\bf Z}_2\times{\bf Z}_2$, that is generated by 
$S_0,S_1,S_2$ (see \cite{Santoprete}).
(The symmetry $S_0$ is the one denoted by $\bar S_0$
in the McGehee coordinates of the previous section.) To obtain certain 
families of periodic solutions, we will use the symmetries $S_0$,$S_1$ 
and $S_2$ in connection with the variational principle according to which
extremum values of the action integral yield periodic solutions of 
the equations (\ref{eqmotion}). To reach this goal we first need to
introduce some notations.

Let $C^\infty([0,T],\mathbb{R}^2)$ be the space of $T$-periodic
$C^\infty$ cycles $f:[0,T]\rightarrow \mathbb{R}^2$. Define 
the inner products
\beq
\begin{array}{l}
\langle f,g \rangle_{L^2}=\int_0^Tf(t)\cdot g(t)dt, \medskip \\
\langle f,g \rangle_{H^1}=\langle f,g\rangle_{L^2}+\langle \dot f, \dot g \rangle_{L^2}, 
\end{array}
\eeq
and let $\|\cdot\|_{L^2}$, $\|\cdot\|_{H^1}$ be the corresponding norms.
Then the completion of $C^\infty([0,T],\mathbb{R}^2)$ with respect to the norm 
$\|\cdot\|_{L^2}$ is denoted by $L^2$ and it is the space of square integrable 
functions. The completion with respect to  $\|\cdot\|_{H^1}$ is denoted by $H^1$ 
and is the Sobolev space of all absolutely continuous $T$-periodic paths that 
have $L^2$ derivatives defined almost everywhere (see \cite{Gordon}).

Let $\Sigma_i([0,T],\mathbb{R}^2)$ denote the subset of $H^1$ formed
by the $S_i$-symmetric paths, with $i\in\{0,1,2,3,4,5,6\}$. It is 
easy to see that each $\Sigma_i$ is a subspace of $H^1$; in fact 
they are Sobolev spaces and have many interesting properties.   
In the following we will restrict our attention to the spaces $\Sigma_i$  
with $i=0,1,2,6$. Let us now prove the following result.
  
\begin{lemma} Let $H^1$ be defined as above, then the subspaces $\Sigma_i$ 
of $S_i$-symmetric paths with $i=0,1,2,6$ are closed, weakly closed, and 
complete with respect to the norm $\|\cdot\|_{H^1}$, and are therefore
Sobolev spaces. Moreover 
\beq
H^1=\Sigma_1\oplus\Sigma_2=\Sigma_0\oplus\Sigma_6.
\eeq
\end{lemma}
{\it Proof:} We first show an interesting fact: we can write $f=(f^1,f^2)$ 
as the sum of an $S_1$ and an $S_2$-symmetric path. Indeed it is well known 
that we can write $f_1$ and $f_2$ as the sum of an even and an odd 
absolutely continuous function, i.e. as $f_1=f_1^e+f_1^o$ and $f_2=f_2^e+f_2^o$.   
Using this idea  we can write the path $f(t)$ as the sum of   
an $S_1$-symmetric function, $f_{S_1}=(f_1^e,f_2^o)$, and an $S_2$-symmetric one,
$f_{S_2}=(f_1^o,f_2^e)$. Now fix an element $f\in  \Sigma_1$. Then
$\langle  f,g \rangle_{H^1}=0$  for every  $g\in \Sigma_2$. This is because
$$\langle  f,g   \rangle_{H^1}=\int_0^T(f_1g_1+f_2g_2)~dt~+~  \int_0^L
(f_1'g_1'+f_2'g_2')~dt,$$ 
where  the first integrand is an  odd function and the second is an odd 
function almost everywhere.  Thus the above scalar product is zero for 
every $g\in\Sigma_2$. 

Let us  denote the space orthogonal to $\Sigma_1$ by
$\Sigma_1^\bot=\{g\in\Sigma_1:\langle  f,g  \rangle_{H^1}=0~ \mbox{for
every}~ g\in \Sigma_1\}$.   It is easy to see  that $\Sigma_1^\bot$ is
closed  and that $S_2\subset  \Sigma_1^\bot$. Now we need to show
that $S_2\supset\Sigma_1^\bot$.  Assume there is $h \in\Sigma_1^\bot$
such that $h\neq 0$ and $h\in\Sigma_2$. Then write $h=h_{S_1}+h_{S_2}$           
and consider $\langle h_{S_1},h_{S_1}+h_{S_2}\rangle_{H^1}$, which means
that $\langle h_{S_1},h_{S_1}\rangle_{H^1}=\|h_{S_1}\|=\delta>0$. But   
this contradicts the hypothesis that $h\in\Sigma_1^\bot$. Therefore
$\Sigma_2=\Sigma_1^\bot$. So $\Sigma_2$ and consequently $\Sigma_1$    
are closed and such  that $H^1=\Sigma_1\oplus\Sigma_2$.  Moreover, since  
$H^1$ is a metric space, $\Sigma_1$ and $\Sigma_2$ are complete. Also 
$\Sigma_1$ and $\Sigma_2$  are weakly  closed  since they  are norm-closed
subspaces. The statements for $\Sigma_0$ and $\Sigma_6$ can be proved in
a similar way. This completes the proof.

\medskip

Let us now introduce some new definitions. We will say that a path in $\Sigma_i$ 
is of class $L_n$, $n=0, \pm1, \pm2, \pm3,\dots$, if its winding number about 
the origin of the coordinate system is $n$ (i.e. if it makes $n$ loops around 
the origin). The sign of $n$ is positive for a counterclockwise rotation and
negative otherwise. Consider the sets $\bar\Sigma_i([0,T],\mathbb{R}^2\backslash\{0\})$. 
Notice that they are open submanifolds of the spaces $\Sigma_i([0,T],\mathbb{R}^2)$ 
and that the family $(L_n)_{n\in{\bf Z}}$ provides a partition of those spaces into homotopy classes, also called components. Two periodic orbits of the isotropic Manev problem ($\mu=1$), one of class $L_8$ and the other of
class $L_{-9}$, are depicted in Fig.~4. 

\begin{figure}[h]
\begin{center}
\resizebox{!}{5cm}{\includegraphics{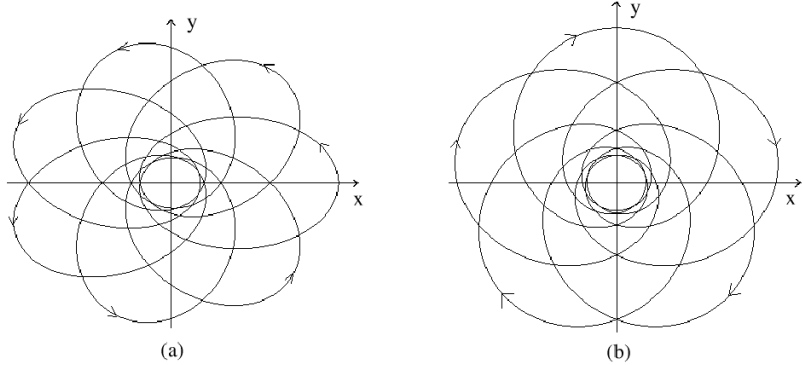}}
\end{center}
\caption {Periodic orbits of the Manev problem:
(a) $S_1$-symmetric periodic orbit of class $L_{8}$, (b) $S_2$-symmetric periodic orbit of class $L_{-9}$. Note that none of these two is $S_3$ symmetric.}
\end{figure}

The Lagrangian  
$L({\bf q}, {\dot {\bf q}})=T({\dot {\bf q}})+U({\bf q})$ of the anisotropic Manev problem given by system (\ref{eqmotion}) has the expression
\beq 
L(x,y,{\dot x},{\dot y})={1 \over  2}({\dot x}^2+{\dot y}^2) + {1 \over  
\sqrt{x^2+ \mu y^2}} + {b \over x^2 +\mu y^2}, \label{L}
\eeq
and the action integral along a path $f$ from time $0$ to time $T$, whose Euclidean coordinate representation is ${\bf q}={\bf q}(t)=(x(t),y(t))$, 
takes the form
$$A_T(f)=\int_0^TL({\bf q}(t), {\bf \dot q}(t))dt.$$

According to Hamilton's principle, the extremals of the functional $A_T$ are solutions of the equations (\ref{eqmotion}). Hence we want to obtain periodic solutions of (\ref{eqmotion}) by finding extremals of the functional $A$.
For this we will use a direct method of the calculus of variation, namely
the lower-semicontinuity method (see \cite{Struwe}). In preparation of a satisfactory theory of existence, the notion of admissible function has to 
be relaxed since the extremals we obtain belong to a Sobolev space. Therefore 
the above method provides only ``weak'' solutions of our problem. To show that 
the paths are regular enough to be classical solutions, we need the following
result, proved in \cite{Gordon}. 

\begin{lemma}
The critical points of $A_T|_{\bar\Sigma_i([0,T],\mathbb{R}^2\backslash \{0\})}$ 
are $T$-periodic solutions of equations (\ref{eqmotion}).
\end{lemma} 

In particular it is well known that if $f$ is a minimizer of the action    
$A_T$ in the space $H^1([0,t],\mathbb{R}^2)$ and if $f$ has no collisions, 
then $f$ is a $T$-periodic solution to (\ref{eqmotion}). Collision have  
to be excluded  because equations (\ref{eqmotion}) break down at collisions  
and because the action is not differentiable at paths with collisions. In 
this paper we are interested to restrict ourself to the spaces $\Sigma_i$ 
of $S_i$-symmetric paths for $i=0,1,2$. The paths that are $S_6$-symmetric 
have to be excluded in the study of periodic orbits since $S_6$-symmetric 
paths must intersect the origin and therefore encounter collisions. 

Now it is not obvious that a collisionless minimizer in $\Sigma_i$ 
is a periodic solution of system (\ref{eqmotion}). However, according 
to the principle of ``symmetric criticality'' (see for example \cite{Chenciner1,Palais}) this is actually true. Indeed,
it can be proved that if $f$ is a collision free path with $dA_t(f)(h)=0$
for every $f\in\Sigma_i$, then $dA_T(f)(h)=0$ for all $f\in H^1([0,T],\mathbb{R}^2)$ and thus $f$ is a critical point in the bigger 
loop space $H^1$ (see \cite{Chenciner1}).

The only obstacle left for applying the direct method is the 
``noncompactness'' of the configuration space. Indeed we want to 
exclude the possibility that the minimizer is obtained when the bodies
are at infinite distance from each other or are collision paths. 
The first problem is solved restricting ourselves to non-simple 
cycles, i.e., to cycles that are not homotopic to a point and thus 
are not in the homotopy class $L_0$. The second problem is solved by 
the following result.

\begin{lemma}
Any family $\Gamma$ of non-simple homotopic cycles in $\bar\Sigma_i([0,T],\mathbb{R}^2\backslash \{0\})$
for $i=0,1,2$ on which $J(f)=\int_0^T{1\over 2}|{\bf\dot q}(t)|^2dt$ and $E(f)=\int_0^T U({\bf q}(t))dt$ are bounded, is bounded away from the origin.
\end{lemma}
The proof of this result follows from \cite{Gordon} if we remark that the anisotropic Manev potential is ``strong'' according to Gordon's definition 
and that the Lagrangian is positive.  

\medskip

To apply the direct method we still need to recall some properties of lower semicontinuous (l.s.c) functions. Let ${\cal F}:X\rightarrow \mathbb{R}$ be 
a real valued function on a topological space $X$. Then $\cal{F}$ is l.s.c. 
if and only if ${\cal F}^{-1}(-\infty,a]$ is closed for every $a\in\mathbb{R}$,
in which case $\cal{F}$ is bounded below and attains its infimum on every 
compact subset of $X$. Moreover when  $X$ is Hausdorff then compact sets are necessarily closed and thus we have the following result. 

\begin{proposition}
Suppose ${\cal F}:X\rightarrow \mathbb{R}$ is a real valued function on an Hausdorff space $X$ and
$${\cal F}^{-1}(-\infty,b]~\mbox{ is compact for every real b}.$$
Then ${\cal F}$ is l.s.c., bounded below, and attains its infimum value on $X$.
\end{proposition}

We can now prove the main result of this section.

\begin{theorem}
For any $T>0$ and any $n=\pm1, \pm2, \pm3,\dots$, there is at least one $S_i$-symmetric ($i=0,1,2$) periodic orbit of the anisotropic Manev problem 
that has period\ \ $T$ and winding number $n$ (i.e., belongs to the homotopy class $L_n$). 
\end{theorem}

{\it Proof:} Let $X$ be a component of $\bar\Sigma_i([0,T],\mathbb{R}^2\backslash \{0\})$ for $i=0,1,2$, that consist of non-simple cycles. Endow $X$ with the 
weak topology it inherits from $\Sigma_i([0,T],\mathbb{R}^2)$. Then $X$ is a subset of an Hilbert space and it is weakly compact if and only if it is weakly closed. 

We wish to apply Proposition~2 with ${\cal F}=A_T$ and thus we have to show that $X \cap A_T^{-1}(-\infty,b]$ is a bounded and weak-closed subset of $\Sigma_i([0,T],\mathbb{R}^2)$.

Since $J=A_T-E$ and $U>0$, we have $E>0$ and therefore
\beq
\begin{split}
&J\leq b\quad \mbox{on}\quad {A_T}^{-1}(-\infty,b]={A_T}^{-1}[0,b],\\
&E=A_T-J\leq b\quad\mbox{on}\quad {A_T}^{-1}[0,b].
\end{split}
\eeq
Since $J\leq b$ the elements of $X$ are bounded in arc length, and from Lemma~5  it follows that the elements of $X$ are bounded away from the origin. Moreover the elements of $X$ are non-simple and thus bounded in the $C^0$ norm and hence in the $L^2$ norm. This last fact combined with $J\leq b$ shows that $X$ is bounded in the $\|\cdot\|_{H^1}$ norm. Thus also $X \cap {A_T}^{-1}(-\infty,b]$ is bounded in the $H^1$ norm.

Now suppose that $\{f_n\}=\{(f^1_n,f^2_n)\}$ is a sequence in $X\cap {A_T}^{-1}[0,b]$ that converges weakly to a cycle
$f\in \Sigma_i([0,T],\mathbb{R}^2)$ for $i=0,1,2$. From general principles, $\|f_n\|_{H^1}$ is bounded and $\|f_n\|_{L^2}\rightarrow \|f_n\|_{L^2}$ 
because weak $\Sigma_i$-convergence implies $C^0$-convergence. 
Since $J(f_n)=1/2\|f_n\|^2_{H^1}-1/2\|f_n\|^2_{L^2}$ it means that   
$J(f_n)$ is bounded and since $E\leq b$ on ${A_T}^{-1}[0,b]$ it follows 
that $\{E(f_n)\}$ is bounded. Moreover, Lemma~5 guarantees
that the functions $f_n$ are bounded away from the origin so that $f$ is homotopic to the $f_n$ in $\mathbb{R}^2\backslash \{0\}$. Therefore $f\in X$. 

To complete the proof we have to show that $f\in {A_T}^{-1}[0,b]$. We know 
that $E(f_n)\rightarrow E(f)$ since weak convergence in $\Sigma_i$ implies 
$C^0$-convergence. For each $n$ let
$$g_n(t)={1\over \sqrt{(f^1_n(t))^2+\mu(f_n^2(t))^2}}+{1\over(f_n^1(t))^2+\mu (f_n^2(t))^2} $$ 
and denote 
$$g(t)={1\over \sqrt{(f^1(t))^2+\mu(f^2(t))^2}}+{1\over(f^1(t))^2+\mu (f^2(t))^2}.$$ 
Each $g_n$ is of class $L^1$ since $A_T(f_n)<\infty$. This implies that the set of all $t$ for which $f_n(t)=0$ has zero measure, otherwise the integral of $g_n(t)$ would be unbounded. So $g_n(t)\rightarrow g(t)$ almost everywhere. 
Also $\int_0^T g_n(t)~dt<A_T(f_n)\leq b$. By Fatou's lemma it follows that 
$g$ is $L^1$ and that
$$ \int_0^T g(t)~dt=\int_0^T\liminf g_n(t) ~dt\leq \liminf \int_0^T g_n(t) ~dt.$$
Now we can use the fact that the norm is weakly sequentially lower semicontinuous (see \cite{Struwe}), thus
$$
\| \dot f\|^2_{L^2}=\|f\|^2_{H^1}-\|f\|^2_{L^2}\leq \liminf \|f_n\|^2_{H^1}-\|f\|^2_{L^2}=
\liminf\|\dot f_n\|^2_{L^2},
$$
where the last equality holds since $\{f_n\}$ converges strongly to $f$ in $L^2$.
Consequently 
\beq
\begin{split}
A_T(f)&={1\over 2}\|\dot f\|^2_{L^2}+\int_0^T g(t)~dt\\
&\leq \liminf{1\over 2}\|\dot f_n\|^2_{L^2}+\liminf\int_0^T g_n(t)~dt
\leq \liminf A_T(f_n)\leq b.
\label{ineq}
\end{split}
\eeq
Relation (\ref{ineq}) now implies that $f\in{A_T}^{-1}[0,b]$.
This completes the proof.

\medskip

Recall now that two intersections of every $S_1$-symmetric ($S_2$-symmetric) 
orbit with the $x$ axis ($y$ axis) must be orthogonal. To distinguish them
from accidental orthogonal intersections, which do not follow because of
the symmetry, we will call them {\it essential orthogonal intersections}.  
From the proof of Theorem~5 and obvious index theory considerations, the 
following result follows 
(see also Fig.~4).

\smallskip

\begin{corollary}
If the essential orthogonal intersections with the $x$-axis ($y$-axis)  
of an $S_1$-symmetric ($S_2$-symmetric) periodic orbit lie on the same 
side of the axis with respect to the origin of the coordinate system, 
then the orbit has an even winding number. If the essential orthogonal intersections are on opposite sides with respect to the origin, then 
the periodic orbit has an odd winding number.
\end{corollary}
 
Since the symmetries $S_0, S_1$ and $S_2$ generate the entire symmetry group, 
it is clear that Theorem~5 captures all periodic orbits with symmetries. This 
result, however, does not tell if other symmetric periodic orbits exist beyond
the the ones with $S_0$, $S_1$ and $S_2$ symmetries. Let us therefore end our paper by proving that $S_3$-periodic orbits do indeed exist. In fact they form 
a rich set if compared to the one of $S_3$-symmetric orbits of the anisotropic Kepler problem (given by (\ref{H}) with $b=0$), which contains only circular orbits. We will show that in our case each homotopy class $L_n$, $n=4k+1$, $k$ integer, contains at least one $S_3$-symmetric periodic orbit. Other homotopy classes may have $S_3$-symmetric periodic orbits, but our approach proves their existence only for winding numbers of the form  $n=4k+1$, $k$ integer.

We consider the set of all paths with one end on the $x$ axis and the other
on the $y$ axis of the coordinate system. As in the case of periodic cycles
discussed in the first part of this section, for a given $T'=T/4>0$ this set 
can be endowed with a Hilbert space structure, the completion of which is 
a Sobolev space. We further divide this space in homotopy classes ${\cal L}_n,
n=0,\pm1,\pm2,\dots$ according to the winding number $n$. 

Using the boundary conditions, it is easy to see that in each class ${\cal L}_n$ the minimizer of the action is a an arc orthogonal to the $x$ and $y$ axes. Its
existence and the fact that it is a solution in the classical sense can be
proved in a similar way as we did for periodic cycles. Once obtaining such
a solution with ends on the $x$ and $y$ axes, we can use the $S_3$-symmetry
and the orthogonality with the axes to complete this solution arc to a periodic 
orbit of period $T>0$. The symmetry implies that the winding number is of the form $n=4k+1$, $k$ integer. This is because if, for example, a solution arc with the 
ends on the $x$ and $y$ axes has a loop around the origin, then the corresponding 
periodic orbit has four loops around the origin. We have thus obtained the following 
result.

\begin{theorem}
For any $T>0$ and any $n=4k+1$, $k$ integer, there is at least one $S_3$-symmetric periodic orbit of the anisotropic Manev problem that 
has period $T$ and winding number $n$ (i.e., belongs to the homotopy 
class $L_n$).
\end{theorem}

It is interesting to note in conclusion that if viewing the anisotropy parameter
as a perturbation and the anisotropic Manev problem as a perturbation of the 
isotropic case (see Section~4), the $S_i$-symmetric ($i=0,1,2$) periodic orbits 
of the isotropic problem are deformed but not destroyed by introducing the 
anisotropy, no matter how large its size. This shows that the $S_i$ symmetries 
($i=0,1,2$) play an important role in understanding the system and are an 
indicator of its robustness relative to perturbations. 

\bigskip

{\bf Acknowledgements.} Florin Diacu was supported in part by the
Pacific Institute for the Mathematical Sciences and by the NSERC 
Grant OGP0122045. Manuele Santoprete was sponsored by a University 
of Victoria Fellowship and a Howard E.\ Petch Research Scholarship.



\begin{thebibliography}{2002}


\bibitem{Burns}
Burns K.\ and  Weiss H.: A  Geometric Criterion  for
Positive Topological  Entropy.\ Commun.\ Math.\ Phys. {\bf  172}, 95-118
(1995).

\bibitem{Chenciner}
Chenciner, A.\ and Montgomery, R.: A remarkable periodic solution of
the three-body problem in the case of equal masses. Ann.\ Math.\ {\bf 152},
881-901 (2000).

\bibitem{Chenciner1}
Chenciner A.: Action minimizing periodic orbits in the Newtonian n-body problem, to appear in the {\it Proceedings of the Evanston Conference dedicated to Don Saari (Dec. 15-19, 1999)}, to appear. 

\bibitem{Gallavotti}
Chierchia, L.\ and Gallavotti G.: Drift and diffusion in phase space.
Ann.\ Inst.\ Henri Poincar\'e, B {\bf 60}, 1-144 (1994).


\bibitem{Chow}
Chow S.-N., Hale J.K., and  Mallet-Paret J.: An Example of Bifurcation to Homoclinic Orbits. J.\ Diff.\ Eqns.\ {\bf 37}, 351-373 (1980).

\bibitem{Chow1}
Chow S.-N.\ and Hale J.K.: {\it Methods of Bifurcation Theory}. New York: Springer Verlag, 1982.

\bibitem{Cicogna0}
Cicogna G.\ and  Santoprete M.: Nonhyperbolic homoclinic chaos. Phys.\ Lett.\ A {\bf 256}, 25-30 (1999).

\bibitem{Cicogna}
Cicogna G.\ and Santoprete M.: An approach to Mel'nikov theory in celestial mechanics. J.\ Math.\ Phys.\ {\bf 41}, 805-815 (2000).

\bibitem{Cicogna1}
Cicogna G.\ and  Santoprete M.: Mel'nikov Method Revisited.
Reg.\ Chaot.\ Dyn.\ {\bf 6}, 377-387 (2001).

\bibitem{Diacu1}
Craig S., Diacu F., Lacomba E.\ A., and Perez E.: On the anisotropic Manev problem.  
J.\ Math.\ Phys.\ {\bf  40}, 1359-1375 (1999).

\bibitem{Delgado}
Delgado, J., Diacu, F., Lacomba, E.A., Mingarelli, A., Mioc, V., 
Perez-Chavela, E., and Stoica, C.: The global flow of the Manev problem, 
J.\ Math.\ Phys.\ {\bf 37} (6) 2748-2761 (1996).

\bibitem{Delshams}
Delshams A.\ and Gutierrez P.: Splitting Potential and the Poincar\'e-Melnikov
Method for Whiskered Tori in Hamiltonian Systems. J.\ Nonlin.\ Sci.\ {\bf 10}, 
435-476 (2000).

\bibitem{Diacu0}
Diacu F., Mingarelli A., Mioc V., and Stoica C.: The Manev two-body problem: quantitative
and qualitative theory. In {\it Dyanamical Systems and Applications}. {\it World Sci.\ 
Ser.\ Appl.\ Anal.\ vol 4}, pp 213-217. River Edge, NJ: World Science Publishing, 1995.

\bibitem{Diacu01}
Diacu, F., Mioc, V., and Stoica, C.: Phase-space structure and regularization
of Manev-type problems. Nonlinear Analysis {\bf 41} 1029-1055 (2000).

\bibitem{Diacu2}
Diacu F.\ and Santoprete M.: Nonintegrability  and chaos in the anisotropic 
Manev problem. Physica D {\bf 156}, 39-52 (2001).

\bibitem{Gordon}
Gordon, W.B.: Conservative Dynamical Systems Involving Strong Forces.
Trans.\ Amer.\ Math. Soc.\ {\bf 204}, 113-135 (1975).

\bibitem{Holmes}
Guckenheimer J.\ and Holmes P.: {\it Nonlinear Oscillations, Dynamical Systems, and Bifurcations of Vector Fields}. New York: Springer, 1983.

\bibitem{Hartman}
Hartman P.: {\it Ordinary Differential Equations}. New York: Wiley, 1969.

\bibitem{Holmes1}
Holmes P.J.\ and Marsden J.E.: Melnikov's method and Arnold diffusion for perturbations 
of Hamiltonian systems. J.\ Math.\ Phys.\ {\bf 23}, 669-675 (1982). 

\bibitem{McGehee}
McGehee R.: Triple collisions in the collinear three-body problem.
Invent.\ Math.\ {\bf 27}, 191-227 (1974).

\bibitem{Melnikov}
Melnikov V.\ K.: On the stability of the center for time periodic perturbations. Trans.\ Moscow Math.\ Soc.\ {\bf 12}, 1-56 (1963).

\bibitem{Palais}
Palais R.: The principle of symmetric criticality. Comm.\ Math\ Phys.\ {\bf 69}, 19-30 (1979).

\bibitem{Poincare}
Poincar\'e, H.: Sur les solutions p\'eriodiques et le principe de moindre action.
Comptes rendus de l'Academie de Sciences, {\bf 123}, 915-918 (1896).

\bibitem{Santoprete}
Santoprete M.: Symmetric periodic solutions of the anisotropic Manev problem. 
J.\ Math.\ Phys.\ {\bf 43}, 3207-3219 (2002).

\bibitem{Struwe}
Struwe M.: {\it Variational Methods. Applications to nonlinear partial differential equations and 
Hamiltonian systems} {\it Second Edition}. Ergebnisse der Mathematik und iherer Grenzgebiete, Berlin: Springer-Verlag, 1996.

\bibitem{Tonelli}
Tonelli L.: Sur une m\'ethode directe du calcul des variations, Rend.\ Circ.\
Matem.\ Palermo {\bf 39}, 233-263 (1915).

\bibitem{Treschev}
Treschev D.: Hyperbolic tori and asymptotic surfaces in Hamiltonian systems. Russ.\ J.\ Math.\ Phys.\ {\bf 2}, 93-110 (1994).

\bibitem{Wiggins}
Wiggins S.: {\it Global Bifurcations and Chaos} New York: Springer, 1988.


\end{thebibliography}
\end{document}